\documentclass[twocolumn]{aastex62}

\shorttitle{Star Formation \& Bars in Interacting Clusters}
\shortauthors{Yoon \& Im}

\begin{document}

\title{Star Formation Enhancement in Barred Disk Galaxies in Interacting Galaxy Clusters}

\correspondingauthor{Myungshin Im}
\email{yyoon@kias.re.kr, mim@astro.snu.ac.kr}

\author[0000-0003-0134-8968]{Yongmin Yoon}
\affiliation{School of Physics, Korea Institute for Advanced Study (KIAS), 85, Hoegiro, Dongdaemun-gu, Seoul, 02455, Republic of Korea}
\affiliation{Center for the Exploration of the Origin of the Universe (CEOU),
Astronomy Program, Department of Physics and Astronomy, Seoul National University, 1 Gwanak-ro, Gwanak-gu, Seoul, 08826, Republic of Korea}

\author{Myungshin Im}
\affiliation{Center for the Exploration of the Origin of the Universe (CEOU),
Astronomy Program, Department of Physics and Astronomy, Seoul National University, 1 Gwanak-ro, Gwanak-gu, Seoul, 08826, Republic of Korea}

\begin{abstract}
A recent study shows that bars can be induced via interaction of galaxy clusters, but it has been unclear if the bar formation by the interaction between clusters is related to the enhancement of star formation. We study galaxies in 105 galaxy clusters at  $0.015<z<0.060$ detected from Sloan Digital Sky Survey data, in order to examine whether the fraction of star-forming galaxies ($f_\mathrm{sf}$) in 16 interacting clusters is enhanced compared with that of the other non-interacting clusters and to investigate the possible connection between the $f_\mathrm{sf}$ enhancement and the bar formation in interacting clusters. We find that $f_\mathrm{sf}$ is moderately higher ($\sim20\%$) in interacting clusters than in non-interacting clusters and that the enhancement of star formation in interacting clusters occurs only in moderate-mass disk-dominated galaxies ($10^{10.0} \le M_\mathrm{star}/M_{\odot} < 10^{10.4}$ and the bulge-to-total light ratio is $\le0.5$). We also find that the enhancement of $f_\mathrm{sf}$ in moderate-mass disk-dominated galaxies in interacting clusters is mostly due to the increase of the number of barred galaxies. Our result suggests that the cluster--cluster interaction can simultaneously induce bars and star formation in disk galaxies.
\end{abstract}


\defcitealias{Yoon2019}{Y19}

\section{Introduction} \label{sec:intro}
Galaxy clusters grow by accretion of galaxies and mergers of galaxy groups and clusters in the $\Lambda$ cold dark matter universe \citep{Berrier2009,McGee2009,Schellenberger2019}. A merger or interaction between galaxy clusters is the most violent event in the universe with kinetic energies up to $\sim10^{64}\,\mathrm{erg}$. Thus, the interacting cluster is a good laboratory to understand how galaxy properties are affected under a violent change of the large-scale environment. For example, \citet[hereafter \citetalias{Yoon2019}]{Yoon2019} recently found observational evidence that cluster--cluster interaction can form bars in disk galaxies, suggesting that such a violent phenomenon is an important mechanism for bar formation. 

One of possible interesting consequences of the cluster--cluster interaction is the enhancement of star formation in the cluster member galaxies. Hence, many studies have been carried out to reveal whether the star formation in galaxies is affected by cluster mergers or interactions. However, the results are somewhat controversial. Several studies based on observation  \citep{Owen1999,Owen2005,Miller2003,Hwang2009,Hou2012,Cohen2014,Cohen2015,Stroe2014,Stroe2017,Sobral2015,Ebeling2019,Soares2019} and simulations \citep{Bekki1999,Bekki2010} show that the star formation is enhanced in merging or interacting clusters. Some of the observational studies \citep{Hou2012,Cohen2014,Cohen2015} found that the fraction of star-forming galaxies is as much as $\sim20$--$30\%$ higher in interacting clusters (clusters with substructures or unrelaxed clusters), and galaxies transform to quiescent ones as the cluster merger process proceeds \citep[e.g.,][]{Cava2017}. These studies suggest various physical mechanisms to enhance star-formation activities of galaxies in interacting clusters: the enhanced time-dependent tidal gravitational field \citep{Bekki1999,Owen2005}, turbulence induced by cluster-wide shock waves in intracluster medium \citep[ICM;][]{Stroe2014,Sobral2015}, and compression of cold gas by increased external pressure of ICM during interaction \citep{Bekki2010}.

On the other hand, a number of studies suggest that star formation is suppressed or not enhanced in interacting clusters \citep{Tomita1996,Fujita1999,Poggianti2004,Chung2009,Haines2009,Shim2011,Tyler2014,Deshev2017,Mansheim2017}. In a recent work, \citet{Okabe2019} examined the fraction of red galaxies with $\log(M_\mathrm{star}/M_{\odot}) > 10.45$ in $\sim180$ merging clusters and $1800$ single clusters from the Hyper Supreme-Cam Subaru Strategic Program. They found that the red fractions are consistent between two cluster classes at $<2\sigma$, although they caution that their result does not exclude the possibility of star formation triggered by cluster interactions. \citet{Fujita1999} argued in their simulation that the increased ram-pressure during cluster--cluster interaction can strip interstellar medium (ISM) of galaxies and thereby suppress star-formation activities. They also showed that the star-formation enhancement by compression of ISM is not significant, which is contrary to the results of \citet{Bekki2010}. \citet{Mansheim2017} suggested that the amplified tidal force and its time variation can remove bound gas in galaxies, which results in suppression of star formation. This is an opposite stand to those of  \citet{Bekki1999} and \citet{Owen2005}.

Another important aspect in interacting clusters is the bar formation by cluster--cluster interactions and how it is related to the star formation in cluster galaxies. For example, \citet{Bekki1999} shows that time-dependent tidal force in interacting clusters can not only trigger star formation but also can contribute to bar formation. In our earlier study \citepalias{Yoon2019}, we have shown that the bar formation can be enhanced by a factor of 1.5 in interacting clusters, backing up the theoretical prediction of \citet{Bekki1999}. 

The bar fraction enhancement in interacting clusters suggests that the bar fraction enhancement might be responsible for the star-formation enhancement in the interacting clusters. Due to the elongated potentials of the structures and materials in bars, it has been suggested that bars can efficiently channel cold gas to the central regions of galaxies and thereby trigger the nuclear star formation in galaxies \citep{Kim2012,Oh2012,Seo2013,Carles2016} or even trigger active galactic nuclei \citep{Oh2012}. Therefore, if the bars are preferentially induced in interacting clusters, one would expect that the star formation in such clusters is also induced in relation to the bar formation.

Motivated by the need for confirming the previously reported star-formation enhancement in interacting clusters and the possible connection between the bar and star formation enhancement in such clusters, we carried out a statistical study using 16 interacting clusters and 89 non-interacting clusters from \citetalias{Yoon2019}. Member galaxies in these clusters are classified as barred or non-barred galaxies, and thus this cluster sample is ideal for investigating the connection between bars and star formation in clusters. In the following, we show that star formation is indeed enhanced in interacting clusters as found in previous studies but the enhancement is dependent on the stellar mass of member galaxies and that the star-formation enhancement is closely related to the bar fraction enhancement in interacting clusters.

 Throughout this paper, we use \emph{H$_0=70$} km s$^{-1}$ Mpc$^{-1}$, $\Omega_{\Lambda}=0.7$, and $\Omega_\mathrm{m}=0.3$ as cosmological parameters, which is supported by observational studies in the past decades \citep[e.g.,][]{Im1997}. 
\\

\begin{figure}
\includegraphics[scale=0.14,angle=00]{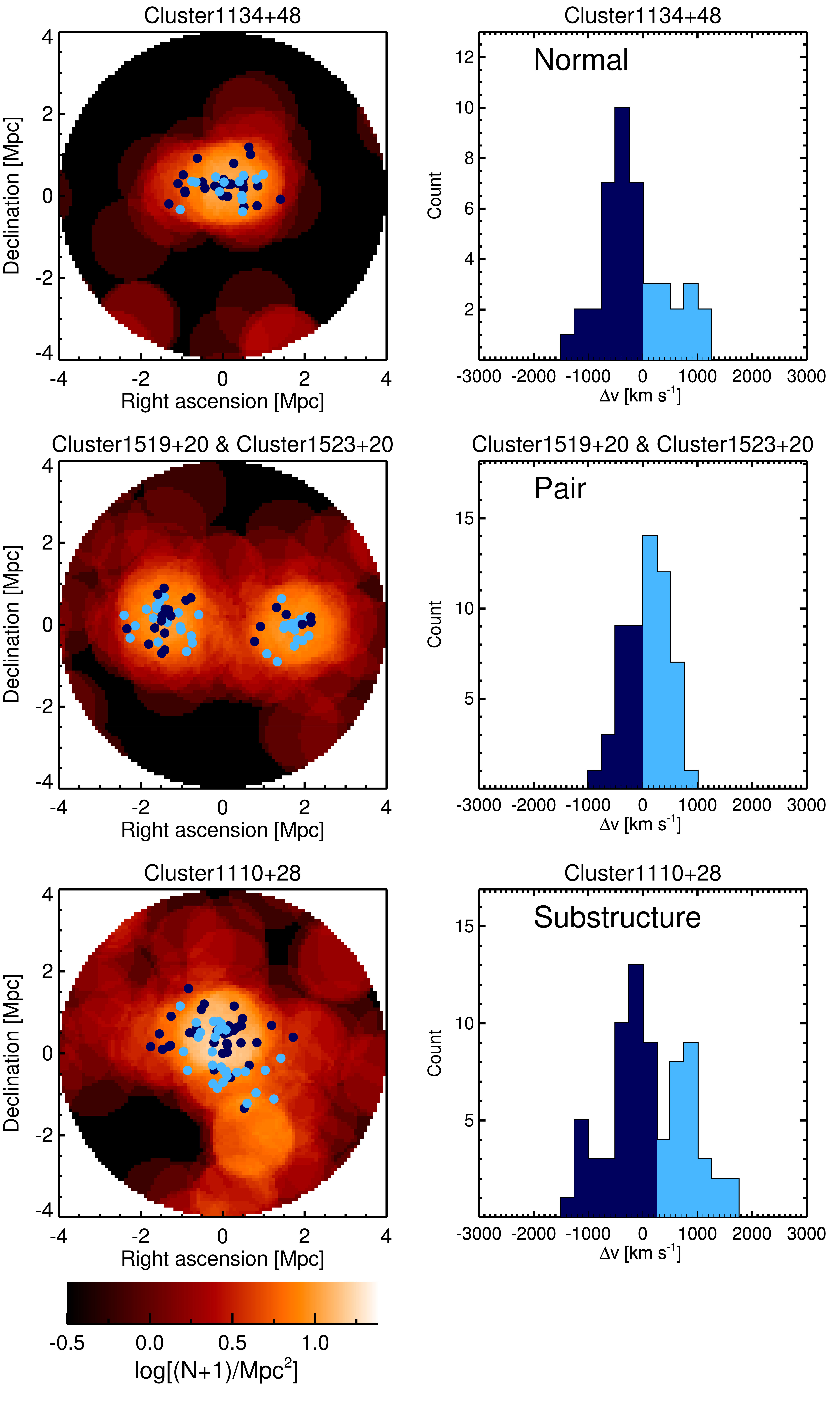}
\centering
\caption{Density maps and radial velocity distributions of three representative types of clusters: undisturbed clusters (top), clusters in close pair (middle), and clusters with substructures (bottom). The left panels show maps of the surface number density of galaxies around clusters and spatial distributions for the cluster member galaxies superimposed onto the maps. The right panels show the velocity distributions for the cluster member galaxies. To construct the density map, we made a grid over a rectangular area of 8 Mpc. Each grid size in the $x$ and $y$ directions was set to be 80 kpc so that total of $100\times100$ points were generated in the grid. At each point, we calculated the surface galaxy number density in an aperture with a radius of 1 Mpc within a rest-frame velocity slice of $\pm2000$ km s$^{-1}$  (see the color bar in the bottom for the color-coded surface number density scale). The member galaxies are split into two groups in the velocity space: one with low redshifts (dark blue) and the other with high redshifts (light blue). We note that in the case of a cluster with substructures (Cluster1110+28), galaxies of the two groups have different spatial distributions.
\label{fig:example}}
\end{figure}

\begin{figure}
\includegraphics[scale=0.34,angle=00]{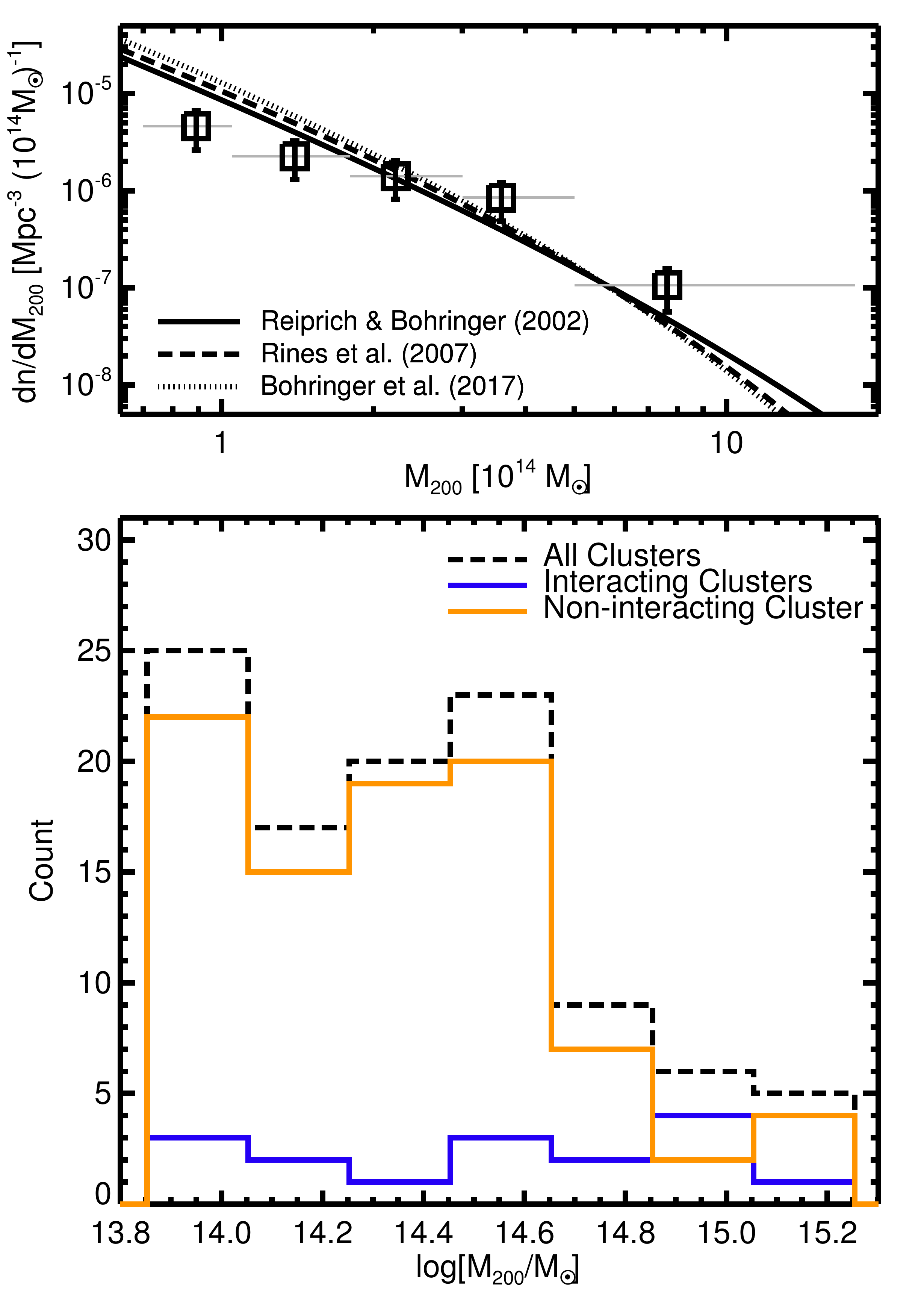}
\caption{Bottom panel shows distributions of $M_{200}$ for all the clusters, interacting clusters, and non-interacting clusters used in this study. The top panel compares the cluster mass function of all clusters with those of previous studies \citep{Reiprich2002,Rines2007,Bohringer2017}. The vertical error bars are the $2\sigma$ Poisson errors of cluster number counts, while the horizontal gray bars indicate the mass ranges. We set the mass ranges in a sense that the similar number of clusters are included in each mass range. We note that the scales of the $x$ axes of the upper and bottom panels are identical.
\label{fig:halodist}}
\end{figure}

\section{Sample and Method} \label{sec:sample}
The samples of galaxies and clusters used in this study are identical to those of \citetalias{Yoon2019}. Details about the samples, cluster identification, selection of interacting clusters, and bar classifications can be found in \citetalias{Yoon2019}. Here, we only briefly describe them.

\subsection{Cluster Identification}
Our sample is based on the MPA--JHU catalog\footnote{\url{http://www.sdss.org/dr14/spectro/galaxy_mpajhu/}} that lists positions and spectroscopic redshifts of galaxies in the Sloan Digital Sky Survey (SDSS) Data Release (DR) 8 \citep{Aihara2011}. This catalog also provides galaxy properties such as stellar masses ($M_\mathrm{star}$) and specific star formation rates (sSFRs: star formation rate per unit $M_\mathrm{star}$). 

Galaxy clusters were selected from a volume-limited sample of galaxies with $\log(M_\mathrm{star}/M_{\odot}) \ge 10.0$ at $0.010 < z < 0.065$. The stellar mass cut ensures that galaxies satisfy the magnitude cut ($m_r<17.77$) of SDSS main galaxy sample \citep{Strauss2002} for spectroscopy. We note that $91\%$ of the spectroscopy target galaxies in the cluster regions (within $R_{200}$) were observed in spectroscopic follow-up,\footnote{The fraction is $92\%$ for targets in all environments.} while this fraction decreases to $86\%$ in the case of the cluster core regions (within $0.2\times R_{200}$) due to the fiber collision between close targets (within $55\arcsec$). Here, $R_{200}$ is a radius within which the mean density is 200 times the critical density of the universe. This lowered fraction of spectroscopically observed galaxies in the cluster core may slightly increase the fraction of star-forming galaxies in clusters, but this should not cause a problem in comparing the star-forming galaxy fractions between clusters of different kinds, since the same spectroscopic target selection bias applies to all the clusters. We also note that there is virtually no difference in the sampling rate for the spectroscopy between the areas of interacting clusters and non-interacting clusters (difference of $\sim1\%$).

The cluster search is performed in the following way. First, we measured the number of galaxies around each galaxy within a projected radius of 1 Mpc and a rest-frame redshift slice of $\delta v = \pm 1000$ km s$^{-1}$ from the galaxy.\footnote{We note that the application of wider velocity slices of 1800 and 2500 km s$^{-1}$ here does not find additional clusters but reduces the number of discovered clusters by $\sim20\%$, especially for those with lower mass. Also see \citet{Lee2019} regarding how the cluster selection can be affected if one chooses a wide velocity cut.} Then, we applied the Friends-of-friends algorithm to connect galaxies in dense environments\footnote{Environments of which surface number densities of galaxies are above 95.4 percentile (or $2\sigma$).} into an overdense region, with a linking length of 1 Mpc in the projected distance and $3000\,$km s$^{-1}$ in the radial velocity.

For the overdense regions, we measured $M_{200}$ that is a cluster mass in $R_{200}$. $M_{200}$ and $R_{200}$ were calculated using all galaxies with $\log(M_\mathrm{star}/M_{\odot}) \ge 10.0$ within a 1 Mpc radius from the center of the overdense region. After excluding non-member galaxies and outliers in the radial velocity space, the one-dimensional (radial) velocity dispersion was derived.  $M_{200}$ and $R_{200}$ were calculated from the one-dimensional velocity dispersions of overdense regions and equations used in the previous studies \citep{Demarco2010,Kim2016}.\footnote{Equations 2 and 3 in \citetalias{Yoon2019}} Through this procedure, we identified 105 galaxy clusters with $M_{200}>7\times10^{13}\,M_{\odot}$ at $0.015 < z < 0.060$.\footnote{Halos with $\sim7\times10^{13}\,M_{\odot}$ can be called groups, rather than clusters. However, for convenience, we define them as clusters in this study.} Note that this redshift range is a bit smaller than the redshift range of the volume-limited galaxy sample to avoid exclusion of member galaxies of clusters near the redshift limits. In total, 4595 galaxies are used in this study, and they are all members of the 105 clusters.\footnote{Member galaxies of a cluster were defined as galaxies within $R_{200}$ from the cluster center and within a rest-frame velocity slice of $\pm3\sigma$, centered on the redshift of the cluster; $\sigma$ is the radial velocity dispersion of the galaxies.}

Among these clusters, we define interacting clusters as clusters that are in close pairs or clusters with substructures in the space and velocity space. The clusters in close pairs are defined as clusters that do not largely overlap within their $R_{200}$, yet that are close enough so that they can be considered to be in a gravitational bound orbit. The conditions can be summarized as (1) the radial velocity difference of the two clusters, $\Delta v$, to be $\Delta v <750$ km s$^{-1}$ (equivalent to $R\lesssim10$ Mpc), and (2) the projected distance between the two clusters 1 and 2, $D$, to be $D<2\times (R_{200,1} + R_{200,2})$, where $R_{200,1}$ and $R_{200,2}$ are $R_{200}$ values of the two clusters in close separation. For more details, see \citetalias{Yoon2019}. We identified clusters with clear substructures using the Dressler--Shectman test \citep{DS88} that finds substructures in clusters by detecting large deviations of local velocity distributions in clusters.\footnote{See Equation 6 in \citetalias{Yoon2019}}

The clusters in a pair are regarded as two individual clusters and hence cluster masses were calculated individually. On the other hand, we regard the cluster with substructure as a single cluster, since substructures are intermingled in the same projected region, which makes it difficult to split the cluster into multiple components. This may lead to overestimation in $M_{200}$ of clusters with substructures. Indeed, three out of five clusters with substructures have $\log(M_{200}/M_{\odot}) > 14.8$ (see Supplementary Table 1 of \citetalias{Yoon2019}).

We found five clusters that have substructures and seven cluster pairs (hence in total 14 clusters are in pairs). Since three clusters belong to both categories, we identified 16 interacting clusters in total. Figure \ref{fig:example} shows examples of the surface number density maps and velocity distributions of galaxies for clusters in isolation, in a pair, and with substructures. The $M_{200}$ distributions for all the clusters, interacting clusters, and non-interacting clusters used in this study are shown in Figure \ref{fig:halodist}. The top panel of Figure \ref{fig:halodist} compares our cluster mass function\footnote{The cluster mass function was calculated by dividing the number of clusters in each bin by the bin size and the comoving volume within the SDSS DR8 Legacy spectroscopic coverage of 7966 square degrees \citep{Aihara2011} and $0.015 < z < 0.060$.} with those of previous studies \citep{Reiprich2002,Rines2007,Bohringer2017}, showing that they are all consistent with each other. The bottom panel shows that the fraction of interacting clusters is higher for higher-mass clusters. This trend is also in the clusters used in \citet{Stroe2017} and \citet{Okabe2019}. This is perhaps due to the $M_{200}$ overestimation for clusters with substructures and/or due to more frequent merging history of massive clusters. The fact that finding substructures could be more efficient for massive clusters with a large number of member galaxies \citep[e.g.,][]{Okabe2019} could also be the reason for the trend.

To check the robustness of the cluster-finding method, we matched the clusters detected here with the Abell clusters \citep{Abell1989} in the SDSS survey area and at $0.02<z<0.055$. The richness parameters (from 0 to 5), indicating how rich the group is in terms of member galaxies, are assigned to the Abell clusters. We find that  $91\%$ (20/22) of the Abell clusters with richness larger than or equal to 1 were detected by our cluster-finding method. This value is the same as the detection rate ($91\%$; 91/100) of mock clusters in GALFORM simulation \citep{Cole2000,Lagos2012}, as found in the test in \citetalias{Yoon2019} that used the same cluster-finding method. For clusters with richness larger than or equal to 2, our method detected all ($100\%$; 5/5). However, only $50\%$ (11/22) of the Abell clusters with the richness of 0 were detected by our method. Therefore, our cluster-finding method is robust in detecting clusters with the Abell richness from 1 to 5.

The definition of interacting or unrelaxed clusters varies between different works, some defining it from the existence of radio-emitting structures \citep{Stroe2017}, symmetry of member galaxy distribution \citep{Cohen2014,Cohen2015,Okabe2019}, and the dynamical state detected from Sunyaev--Zeldovich effect or X-ray emission \citep{Rossetti2016}. Although the fraction of unrelaxed or interacting clusters is found to be $\sim10\%$ in \citet{Okabe2019}, some works find $40\%$, significantly more than what we found here. This discrepancy must be due to how interacting clusters are defined. If we loosen our criteria for selecting interacting clusters, using the (projected and radial) pair separation distance that is 1.25 (or 1.5) times larger than what we originally adopted and the probability of having substructures above $95\%$ (or $90\%$) instead of $99.99\%$, the number of interacting clusters is 41 (or 52). We note that the use of the loosened definition does not reverse the basic results on the $f_\mathrm{sf}$ difference between interacting and non-interacting clusters in Section \ref{sec:results}, although the difference is reduced from a factor of 1.2 to 1.1.
\\

\subsection{Bar Classification, $B/T$, and sSFR of Galaxies}

  Bars were identified through a quantitative method using the IRAF ELLIPSE \citep{Jedrzejewski1987}, augmented by visual classification. For the bar classification, we used galaxies that have ellipticities less than or equal to 0.5, which corresponds to an inclination angle smaller than or equal to $60^{\circ}$. This is because it is difficult to detect bars in highly inclined galaxies. We detected bars with several quantitative criteria that find an elongated structure (high ellipticity) for several consecutive ellipses, but an abrupt drop of the ellipticity at the end of the elongated structure, and an almost constant position angle over the high-ellipticity region. We visually inspected all galaxies to complement the automated bar classification. By doing so, we excluded false detections and added apparent bar galaxies that were missed in the automated classification. 
  
We derived the ($r$-band) bulge-to-total light ratio ($B/T$) of each galaxy based on radial surface brightness profiles that were extracted by the ELLIPSE and using a model that is combination of the de Vaucouleurs law for bulges and the exponential profile for disks. The $B/T$ value traces how a bulge is dominant in a galaxy total luminosity. Thus, $B/T$ values are usually used to quantitatively define the morphology of galaxies: generally, $B/T\lesssim0.5$ corresponds to disk-dominated late-type galaxies, while $B/T\gtrsim0.5$ indicates bulge-dominated early-type galaxies \citep{Fukugita1998,Im2002,Oohama2009}.
  
\begin{figure*}
\includegraphics[scale=0.275,angle=00]{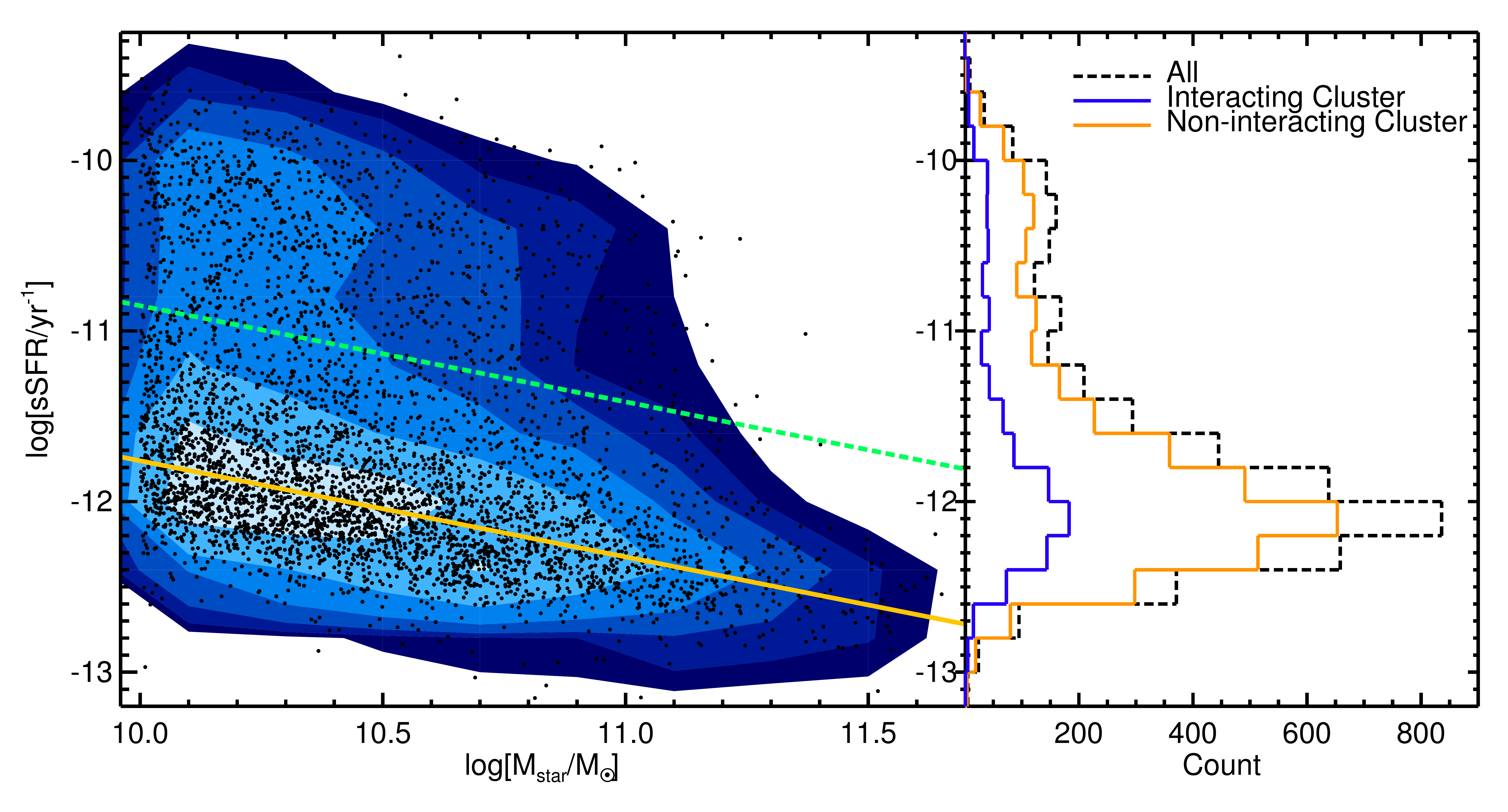}
\centering
\caption{Distribution of all galaxies in the $M_\mathrm{star}$ and sSFR plane (the left panel).  The yellow solid line indicates the low-sSFR sequence. The green dashed line denotes the dividing line between high- and low-sSFR galaxies (Equation \ref{eq:divide}). We also show sSFR distributions for interacting clusters, non-interacting clusters, and all clusters (the right panel).
\label{fig:ssfrdist}}
\end{figure*}

We used galaxy sSFRs from the MPA--JHU catalog. In the catalog, SFRs were derived from nebular emission lines. In addition, galaxy photometry is also used to compute SFRs outside of fibers. In the case of active galactic nuclei and galaxies with weak emission lines, SFRs were derived from photometry.\footnote{We find that the linear Pearson correlation coefficient between sSFRs and $u-r$ color values of galaxies is $-0.88\pm0.01$. The sSFR values have a strong anticorrelation with $u-r$ color, which is a well-known proxy for young stellar populations. We note that use of $u-r$ color instead of sSFRs does not change our main results. }

Because 34 among 4595 galaxies do not have sSFR information, we assigned sSFRs derived from spectral energy distribution fitting to these galaxies. For this purpose, we utilized the code Fitting and Assessment of Synthetic Templates \citep[FAST;\footnote{\url{http://w.astro.berkeley.edu/~mariska/FAST.html}}][]{Kriek2009}, which performs $\chi^2$ fitting of the broadband photometry ($u, g, r, i,$ and $z$ in this study) and works with stellar population grids to derive the best-fit model and its properties.\footnote{More specifically, we used the initial mass function of \citet{Chabrier2003} and assumed a delayed exponentially decreasing SFR. We modeled the stellar population with the $e$-folding time scales, $8.9\leq$ $\log$($\tau$/yr)  $\leq11.0$ with a step size of 0.1 and the ages of $9.5\leq$ $\log$($t$/yr) $\leq10.0$ with a step size of 0.1. We used several metallicity values (Z = 0.004, 0.008, 0.02, and 0.05). For dust attenuation model, we used the attenuation law from \citet{Calzetti2000}. We adopted the extinction values at $V$ band in $0.0\leq$ $A_{V}$  $\leq3.0$ with a step size of 0.1.}

Figure \ref{fig:ssfrdist} shows the distribution of all galaxies in the $M_\mathrm{star}$ and sSFR plane. Low-sSFR galaxies are clustered at $\log(\mathrm{sSFR}/\mathrm{yr}^{-1})<-11.0$ but with a gradient in the $M_\mathrm{star}$--sSFR plane. Therefore, to divide galaxies into high- and low-sSFR ones, rather than adopting a single sSFR cut, we fit the low-sSFR sequence and divide galaxies into high- and low-sSFR ones. To do so, we conducted a linear fit using robust least absolute deviation\footnote{IDL LADFIT procedure} to galaxies with $\log(\mathrm{sSFR}/\mathrm{yr}^{-1})<-11.0$. During the process, we obtained the mean of the absolute deviation (in a logarithmic value) between the linear relation and sSFR values (hereafter, MAD). Then, galaxies that have sSFRs higher than $3.5 \times \mathrm{MAD}$ above the linear relation were excluded. With the remaining galaxies, we repeated the above procedure until the linear fit result converged. The converged linear relation represents the low-sSFR sequence, indicated by the yellow solid line in Figure \ref{fig:ssfrdist}. The converged MAD is 0.26 dex. Finally, we set a dividing line between high- and low-sSFR galaxies at $3.5 \times \mathrm{MAD}$ (i.e., 0.91 dex) above the low-sSFR sequence relation, denoted by the green dashed line in Figure \ref{fig:ssfrdist}. The equation of this dividing line is
\begin{equation}
\log(\mathrm{sSFR}/\mathrm{yr}^{-1})=-0.56 \log(M_\mathrm{star}/M_{\odot}) - 5.22.
\label{eq:divide}
\end{equation}
We note that use of a simple cut of $\log(\mathrm{sSFR}/\mathrm{yr}^{-1})=-11.0$ instead of the dividing line for separation does not change the results presented in Section \ref{sec:results}, which means our results are insensitive to minor changes in the specific definition for high- and low-sSFR galaxies.

In this study, we divide galaxies into several categories according to sSFR, $M_\mathrm{star}$, $B/T$, and presence of bar. In Table \ref{tb:num}, we show the total numbers of galaxies in each category.
\\

\begin{deluxetable*}{rcc}
\tablecaption{Total Number of Galaxies in Each Category\label{tb:num}}
\tabletypesize{\scriptsize}
\tablehead{\colhead{Category} & \colhead{Non-interacting Clusters} & \colhead{Interacting Clusters}
}
\startdata
All&3577&1018\\
\hline
$B/T\le0.5$ and $\log(M_\mathrm{star}/M_{\odot}) < 10.4$& 903& 272\\
$B/T\le0.5$ and $\log(M_\mathrm{star}/M_{\odot}) \ge 10.4$& 803& 251\\
$B/T>0.5$ and $\log(M_\mathrm{star}/M_{\odot}) < 10.4$& 691& 195\\
$B/T>0.5$ and $\log(M_\mathrm{star}/M_{\odot}) \ge 10.4$&1072& 285\\
\hline
Barred, $B/T\le0.5$ and $\log(M_\mathrm{star}/M_{\odot}) < 10.4$& 130&  53\\
Non-barred, $B/T\le0.5$ and $\log(M_\mathrm{star}/M_{\odot}) < 10.4$& 457& 111\\
Barred, $B/T\le0.5$ and $\log(M_\mathrm{star}/M_{\odot}) \ge 10.4$& 158&  79\\
Non-barred, $B/T\le0.5$ and $\log(M_\mathrm{star}/M_{\odot}) \ge 10.4$& 314&  75\\
\enddata
\tablecomments{Among all 4595 galaxies, 123 galaxies do not have $B/T$ information. They are too compact to constrain $B/T$ or edge-on galaxies. See \citetalias{Yoon2019} for more information. For the categories in which bars were classified, we used galaxies that have ellipticities less than or equal to $0.5$ to avoid galaxies with large inclination angles that are difficult to detect bars.
}
\end{deluxetable*}

\section{Results} \label{sec:results}

\begin{figure*}
\includegraphics[scale=0.25,angle=00]{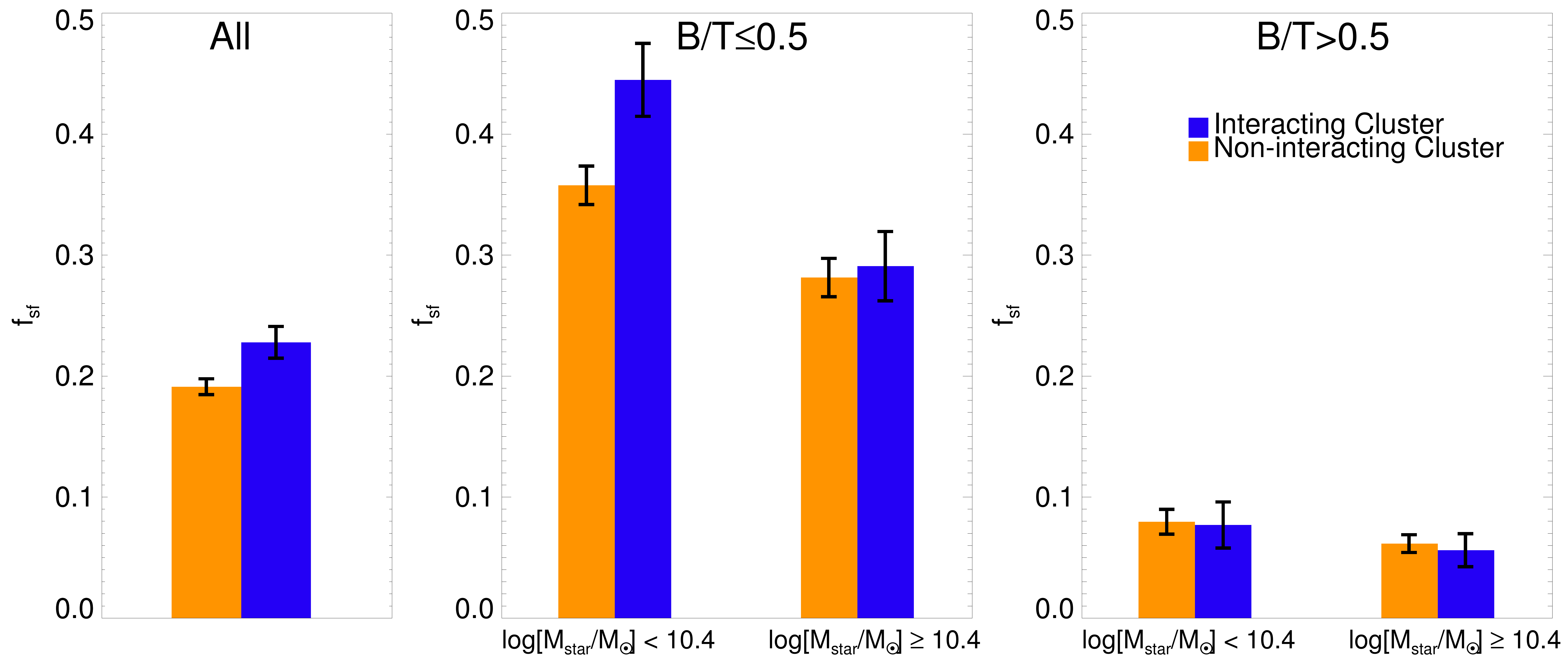}  
\centering
\caption{Fractions of high-sSFR galaxies ($f_\mathrm{sf}$) in interacting clusters and non-interacting clusters. The left panel is for all galaxies. The middle and the right panels are for galaxies with $B/T\le0.5$ and $B/T>0.5$, respectively. In each $B/T$ bin, we divided galaxies into two mass bins of $\log(M_\mathrm{star}/M_{\odot}) < 10.4$ and $\log(M_\mathrm{star}/M_{\odot}) \ge 10.4$. 
\label{fig:sffrac}}
\end{figure*}

 \begin{figure}
\includegraphics[scale=0.29,angle=00]{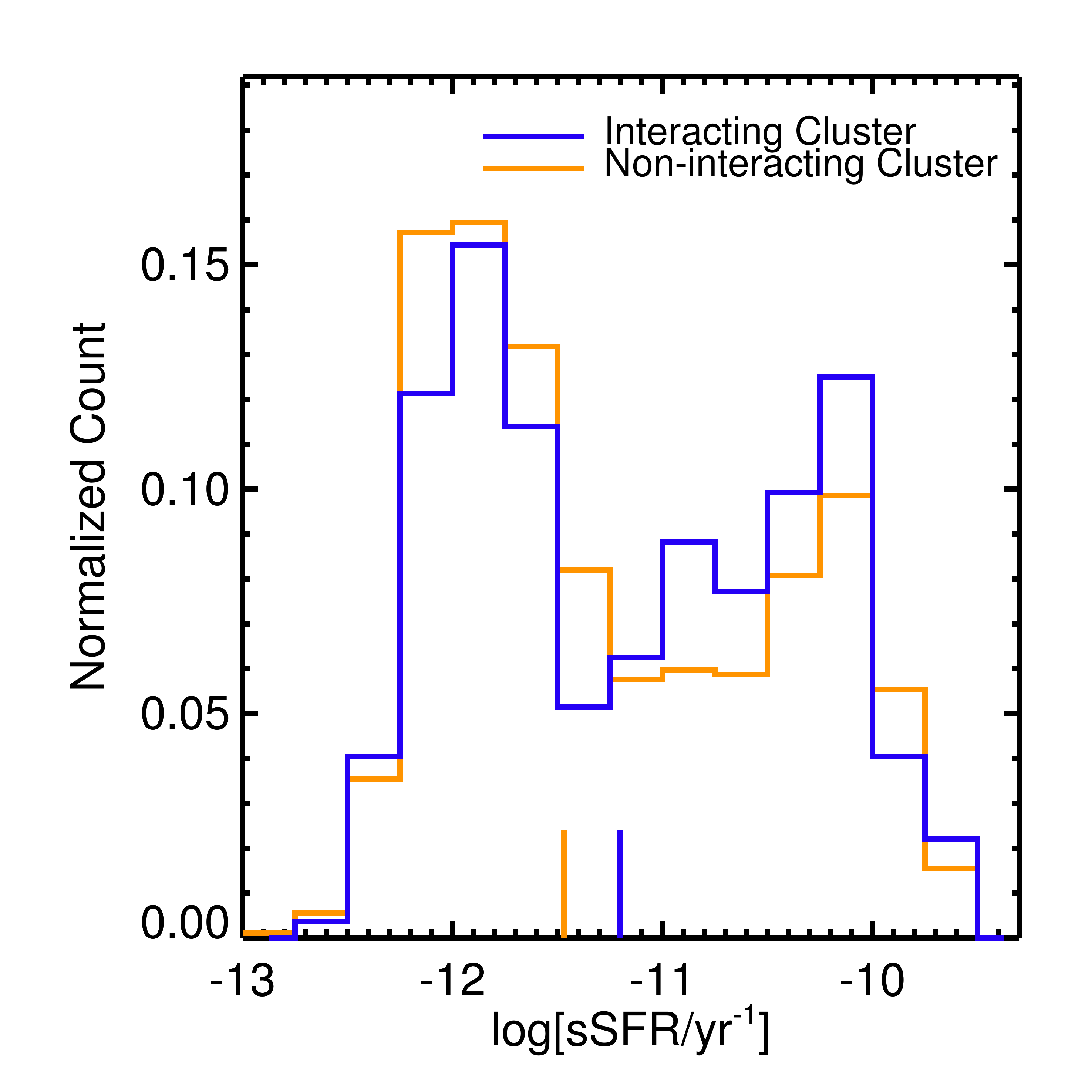}  
\centering
\caption{sSFR distributions for galaxies with $B/T\le0.5$ and $\log(M_\mathrm{star}/M_{\odot}) < 10.4$ in interacting clusters and non-interacting clusters. The vertical lines denote median values of the distributions. The total count of each distribution is normalized to unity.
\label{fig:indgal}}
\end{figure}

We calculated fractions of high-sSFR galaxies ($f_\mathrm{sf}$) of interacting and non-interacting clusters. $f_\mathrm{sf}$ is defined as $N_\mathrm{high}/N_\mathrm{all}$, where $N_\mathrm{high}$ is the number of galaxies located above the sSFR dividing line (Equation \ref{eq:divide}) in Figure \ref{fig:ssfrdist} and $N_\mathrm{all}$ is the number of all galaxies (high- and low-sSFR galaxies). The error for the fraction in this study is the standard error for the proportion for a binomial distribution (see Equation 8 in \citetalias{Yoon2019}). 

 As shown in the left panel of Figure \ref{fig:sffrac}, we find that $f_\mathrm{sf}$ is $1.19\pm0.08$ times higher in interacting clusters than in non-interacting clusters ($0.228\pm0.013$ versus $0.191\pm0.007$). In order to find out which galaxies account for the difference in $f_\mathrm{sf}$, we divided galaxies into four categories on the basis of their $M_\mathrm{star}$ and $B/T$ values: $\log(M_\mathrm{star}/M_{\odot}) < 10.4$ (moderate mass), $\log(M_\mathrm{star}/M_{\odot}) \ge 10.4$ (high mass),\footnote{We can clearly see a different trend, when galaxies are segregated at $\log(M_\mathrm{star}/M_{\odot}) = 10.4$, which is also close to the median $M_\mathrm{star}$ of 4595 galaxies. In this paper, we use the expression ``moderate-mass galaxies'' instead of ``low-mass galaxies'' as a counterpart of the high-mass galaxies, since the stellar masses of these galaxies ($10.0\le \log(M_\mathrm{star}/M_{\odot}) < 10.4$) are barely below $M^*$, which is $\sim10^{10.6}\,M_{\odot}$ \citep{Kelvin2014}.} $B/T>0.5$ (bulge dominated), and $B/T\le0.5$ (disk dominated). \citetalias{Yoon2019} show that the bar fraction enhancement in interacting clusters is significant at $B/T\le0.5$. Thus, \citetalias{Yoon2019} used the criterion of $B/T\le0.5$ to define disk-dominated galaxies for further investigation. Here, we also used the same criterion as in \citetalias{Yoon2019} to define disk-dominated galaxies for the purpose of examining correlation between the enhancement of bar fraction in interacting clusters and star formation. Minor adjustments of this criterion (e.g., $B/T\le0.3$--$0.7$) do not essentially change our main conclusion.
 
  The middle and right panels of Figure \ref{fig:sffrac} show $f_\mathrm{sf}$ of each category. We find that $f_\mathrm{sf}$ of bulge-dominated galaxies in non-interacting clusters is comparable to that of interacting clusters within the error, for both mass ranges. $f_\mathrm{sf}$ of high-mass disk-dominated galaxies in non-interacting clusters is also similar to that of interacting clusters within the error. However, in the case of moderate-mass disk-dominated galaxies only, $f_\mathrm{sf}$ of the non-interacting clusters is different from that of interacting clusters, in such a way that $f_\mathrm{sf}$ of interacting clusters is $1.24\pm0.10$ times higher than that of non-interacting clusters ($0.445\pm0.030$ versus $0.358\pm0.016$). This is confirmed by sSFR distributions for moderate-mass disk-dominated galaxies in interacting clusters and non-interacting clusters shown in Figure \ref{fig:indgal}. The probability $(0\le P \le1)$ of the null hypothesis that the two distributions in the figure are drawn from the same distribution is 0.044 by the Kolmogrov--Smirnov test. We note that the $f_\mathrm{sf}$ increase in moderate-mass disk-dominated galaxies is responsible for $\sim90\%$ of the total $f_\mathrm{sf}$ increase in interacting clusters compared to non-interacting clusters.
  
 \begin{figure}
\includegraphics[scale=0.29,angle=00]{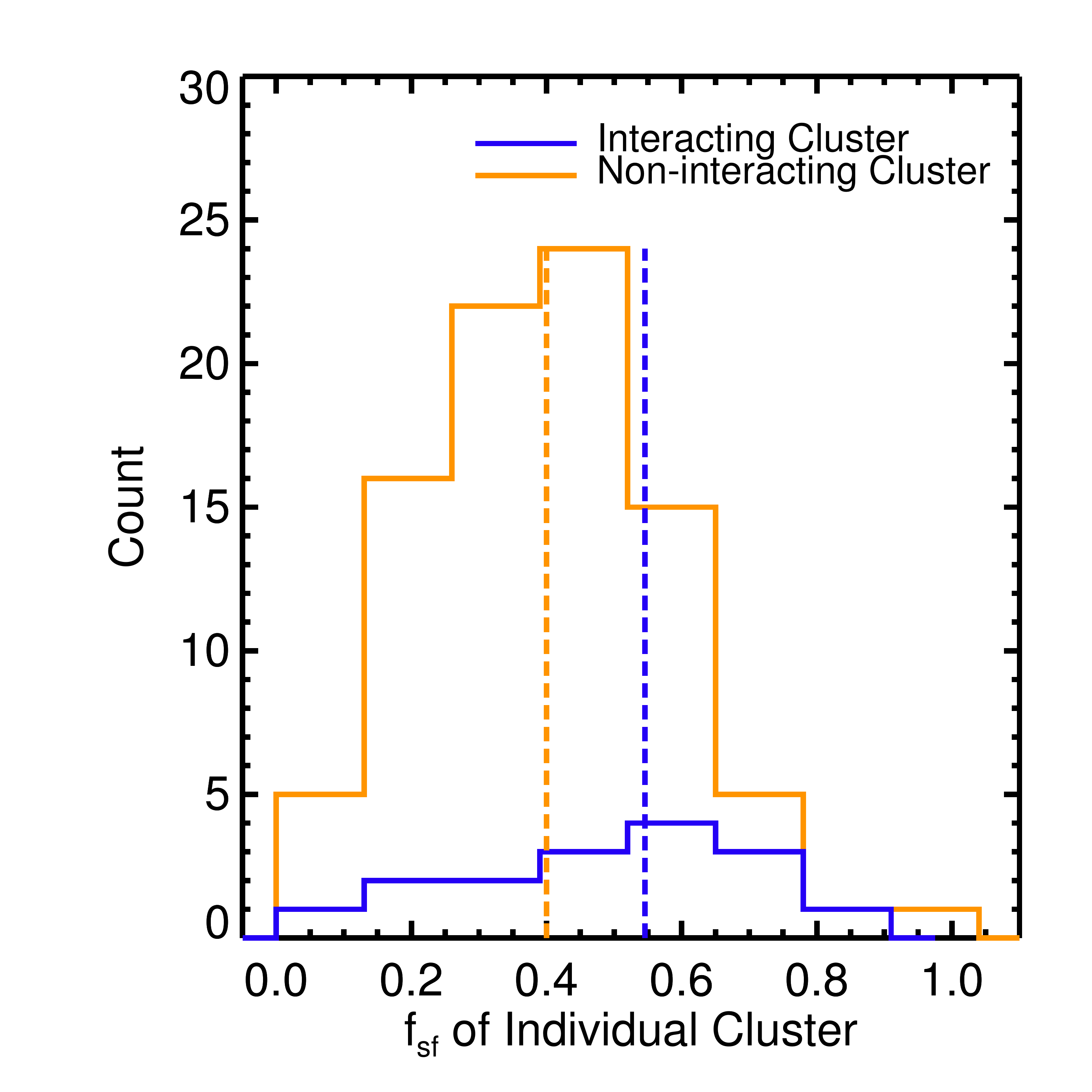}  
\centering
\caption{Distributions for $f_\mathrm{sf}$ of individual clusters in which only galaxies with $B/T\le0.5$ and $\log(M_\mathrm{star}/M_{\odot}) < 10.4$ are used. The vertical dashed lines indicate median values of the distributions.
\label{fig:sffrac2}}
\end{figure} 
  
  A similar result is derived from distributions for $f_\mathrm{sf}$ of individual clusters in which only moderate-mass disk-dominated galaxies are used. The distributions are shown in Figure \ref{fig:sffrac2}. In the figure, the distribution for interacting clusters is skewed to higher $f_\mathrm{sf}$ compared to that of non-interacting clusters. The probability $(0\le P \le1)$ of the null hypothesis that the two distributions are drawn from the same distribution is 0.105 based on the Kolmogrov--Smirnov test.

\begin{figure*}
\includegraphics[scale=0.48,angle=00]{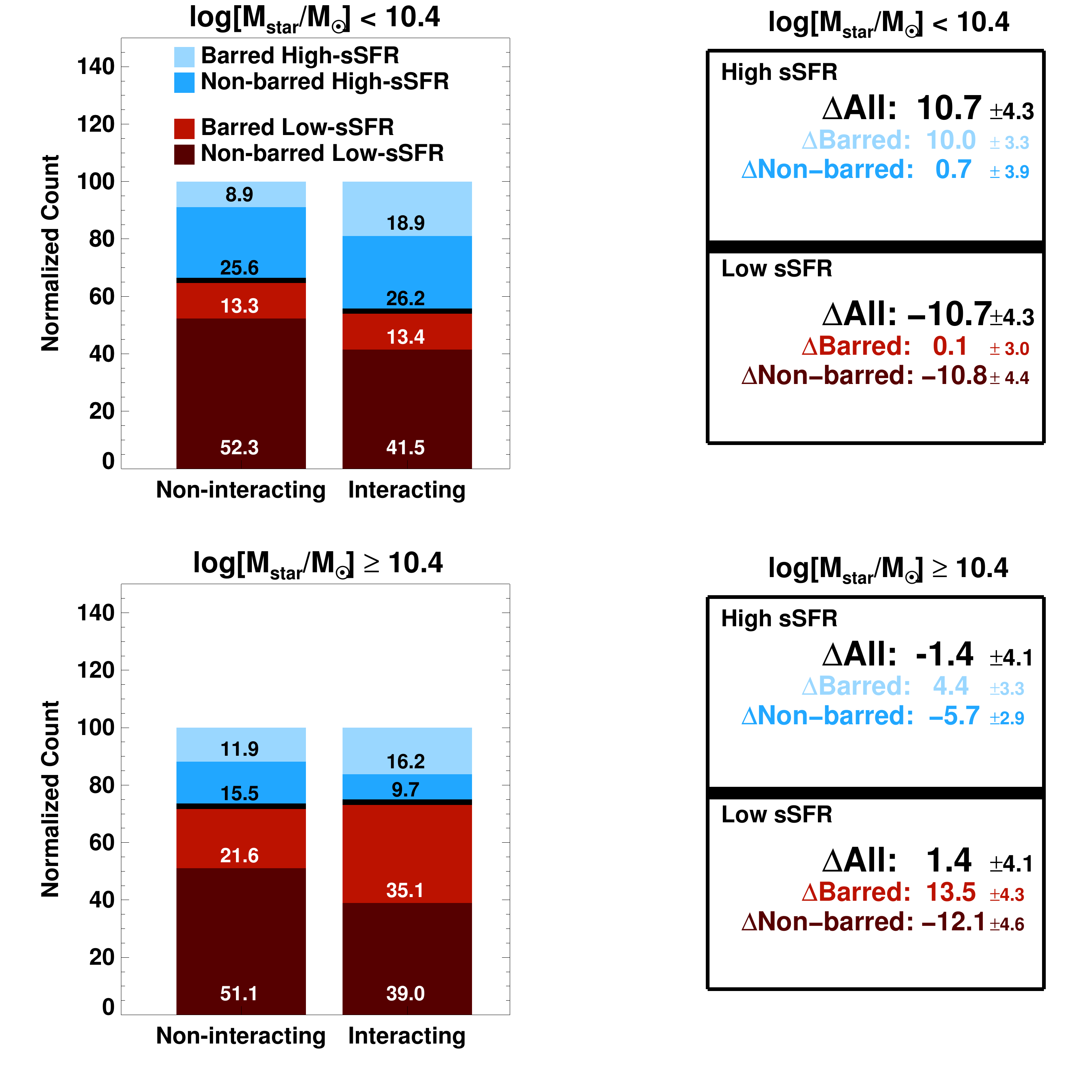}  
\centering
\caption{Proportions of four populations of galaxies with $B/T\le0.5$ (barred high-sSFR, non-barred high-sSFR, barred low-sSFR, and non-barred low-sSFR galaxies) depicted in bar charts (the left panels), and differences of the normalized counts of each population (the right panels), which are calculated by subtracting the normalized counts of non-interacting clusters from those of interacting clusters. Disk-dominated galaxies were divided into two $M_\mathrm{star}$ bins: $\log(M_\mathrm{star}/M_{\odot}) < 10.4$ (the upper panels) and $\log(M_\mathrm{star}/M_{\odot}) \ge 10.4$ (the lower panels). In a mass bin, we normalized the total number of disk-dominated galaxies in non-interacting clusters (or interacting clusters) to 100.  The errors of the differences were computed by random sampling (100,000 times).
\label{fig:num}}
\end{figure*}

To understand the interplay between bars in galaxies and their sSFRs, we investigated changes in proportions of four populations of disk-dominated galaxies (barred high-sSFR, non-barred high-sSFR, barred low-sSFR, and non-barred low-sSFR galaxies) in each mass bin (moderate-mass or high-mass bin).\footnote{As mentioned in Section \ref{sec:sample}, we used galaxies with $e\le0.5$ hereafter.} The results are shown in Figure \ref{fig:num} as bar charts and differences of the normalized counts of each population. The difference in the normalized counts were derived by subtracting the normalized counts of non-interacting clusters from those of interacting clusters. 

 In the case of $\log(M_\mathrm{star}/M_{\odot}) < 10.4$, the fraction of non-barred low-sSFR galaxies in interacting clusters decreases by $10.8\pm4.4\%$ point in comparison with that of non-interacting clusters. Meanwhile, almost the same amount of barred high-sSFR galaxies ($10.0\pm3.3\%$) increases. However, the barred low-sSFR and non-barred high-sSFR galaxies show negligible changes in the proportion within the errors. The net effect is that $f_\mathrm{sf}$ in interacting clusters increases by $10.7\pm4.3\%$ point (or a factor of $\sim1.3$) compared with that of non-interacting clusters, which is consistent with the middle panel of Figure \ref{fig:sffrac}. Another net effect is the enhancement of bar fraction ($1.45$ times) in interacting clusters as in \citetalias{Yoon2019}. Although many different paths can be considered for the change in $f_\mathrm{sf}$ of galaxies in each category, the most straightforward interpretation is that the enhancement of $f_\mathrm{sf}$ in interacting clusters is almost entirely due to the transformation of non-barred low-sSFR galaxies into barred high-sSFR galaxies. This implies that the star-formation enhancement in moderate-mass disk-dominated galaxies of interacting clusters is related to the bar formation via cluster--cluster interactions. 

In the case of $\log(M_\mathrm{star}/M_{\odot}) \ge 10.4$, the transition between low-sSFR galaxies and high-sSFR galaxies is negligible ($1.4\pm4.1\%$ point), when comparing interacting clusters to non-interacting clusters, which is also in agreement with the middle panel of Figure \ref{fig:sffrac}. However, the fraction of barred galaxies in interacting clusters increases by a factor of $\sim1.5$ within each low- or high-sSFR bin as in \citetalias{Yoon2019}. Adopting the most simplistic interpretation again, this result suggests that the bar formation by cluster--cluster interaction is not noticeably related to star-formation enhancement in high-mass disk-dominated galaxies, which is in contrast with their moderate-mass counterparts.
\\

\begin{figure}
\includegraphics[scale=0.14,angle=00]{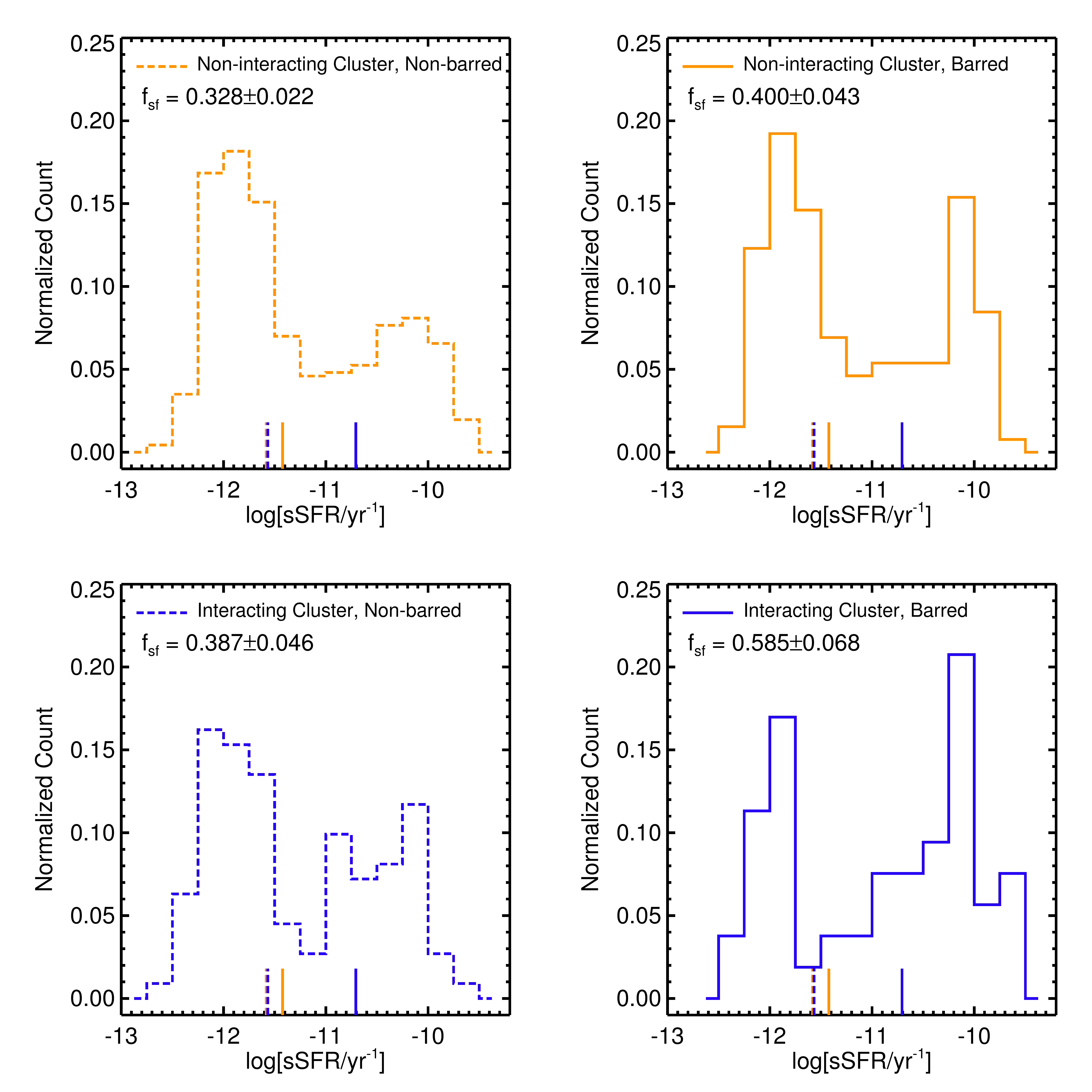}  
\centering
\caption{sSFR distributions for galaxies with $B/T\le0.5$ and $\log(M_\mathrm{star}/M_{\odot}) < 10.4$ in interacting clusters and non-interacting clusters. We divided the moderate-mass disk-dominated galaxies into barred and non-barred galaxies. The vertical lines correspond to median sSFRs of the four distributions. In each panel, we show $f_\mathrm{sf}$ of galaxies in each category.
\label{fig:4p}}
\end{figure}

\section{Discussion} \label{sec:discuss}
In the previous section, we show that $f_\mathrm{sf}$ is moderately enhanced in interacting clusters compared with non-interacting clusters, and the enhancement occurs only in moderate-mass disk-dominated galaxies. This is in contrast to the bar fraction enhancement that was found to occur more in high-mass disk-dominated galaxies in interacting clusters \citepalias{Yoon2019}. Furthermore, the enhancement of $f_\mathrm{sf}$ in moderate-mass disk-dominated galaxies is found to be directly related to the increase in the number of galaxies with bars. 

Our result on the enhancement of $f_\mathrm{sf}$ in interacting clusters is comparable to the results found in \citet{Cohen2014,Cohen2015}, who used SDSS data and a large number (over 100) of clusters. They found that $f_\mathrm{sf}$ is $\sim20$--$30\%$ higher in clusters with substructures or unrelaxed clusters, which is similar to or slightly higher than the value of our result ($20\%$ or up to $24\%$ depending on the galaxy types). The result of \citet{Cohen2014} is mainly based on galaxies brighter than $M_r=-20.5$. Borrowing the $M_r$--$M_\mathrm{star}$ conversion formula\footnote{$\log(M_\mathrm{star}/M_{\odot})=-0.39M_r+1.05(g-r)+1.60$} in \citet{Yoon2017}, $M_r=-20.5$ corresponds to $\log(M_\mathrm{star}/M_{\odot})\sim10.0$ for star-forming galaxies and  $\log(M_\mathrm{star}/M_{\odot})\sim10.4$ for quiescent galaxies. Therefore, their magnitude-cut sample is similar to our mass-cut sample, except that they miss moderate-mass quiescent galaxies in our sample, and hence their result is consistent with ours.

Our result is also similar to the result of \citet{Hou2012}, in which they used 15 rich groups and reported the enhancement of $f_\mathrm{sf}$ ($\sim28\%$) in groups with substructures. \citet{Stroe2017} used 19 clusters with a total sample of over 3000 galaxies and found that the H$\alpha$ luminosity function for clusters in mergers shows a higher characteristic density than relaxed clusters, which qualitatively agrees with our result. Overall, our finding of the $f_\mathrm{sf}$ enhancement in interacting clusters confirms results from previous works.

On the other hand, we find an unique aspect of the $f_\mathrm{sf}$ enhancement in that the $f_\mathrm{sf}$ enhancement is mostly from moderate-mass galaxies with $10.0 \le \log(M_\mathrm{star}/M_{\odot}) < 10.4$. Note that \citet{Okabe2019} did not find the $f_\mathrm{sf}$ enhancement (within $2\sigma$) between merging and single clusters for galaxies with $\log(M_\mathrm{star}/M_{\odot}) > 10.45$, although they did not completely reject the possibility of star-formation enhancement in merging clusters. Their result can be understood as a result of the mass dependence of the $f_\mathrm{sf}$ enhancement.

The $M_\mathrm{star}$-dependent trend in $f_\mathrm{sf}$ can be explained by the different amount of gas in disk galaxies with different $M_\mathrm{star}$. It is known that lower-mass disk galaxies have more plentiful gas than higher-mass disk galaxies \citep{Erb2006,Hopkins2009,Masters2012}. According to this notion, for moderate-mass disk-dominated galaxies, the bar formation in interacting clusters can be easily accompanied with the SFR enhancement, since they have a relatively high amount of gas.\footnote{Several previous studies show that some galaxies in clusters in merging processes or with substructures can have large amounts of \ion{H}{1} gas or molecular gas contents that are comparable to field galaxies \citep{Stroe2015,Cybulski2016,Cairns2019}.} On the other hand, the gas contents are less abundant for high-mass disk-dominated galaxies. Thus, bars can be triggered easily in those galaxies as shown in \citetalias{Yoon2019} and Figure \ref{fig:num}, since the less abundant gas contents for a given $M_\mathrm{star}$ in disk galaxies are more favorable for formation of bars \citep{Berentzen1998,Berentzen2004,Bournaud2005,Villa-Vargas2010,Masters2012}.  However, due to the low amount of gas, the formation of bars in interacting clusters is not translated into the triggering of star formation in high-mass disk-dominated galaxies.

In \citetalias{Yoon2019}, we argued that the time-dependent tidal gravitational field during the cluster--cluster interaction is responsible for inducing bars as suggested by a simulation work of \citet{Bekki1999}. The enhancement of $f_\mathrm{sf}$ in moderate-mass disk-dominated galaxies can be understood under the same framework. Specifically, the time-dependent tidal force in interacting clusters exerts non-axisymmetric perturbation to a disk galaxy and subsequently creates a bar. Then, the bar structure exerts forces onto gas components and makes the gas funnel into the central region of the galaxy, thereby triggering the star formation there \citep{Kim2012,Seo2013,Carles2016}. In this manner, galaxies with the newly formed bars in interacting clusters can also become the newly triggered high-sSFR galaxies. Indeed, as shown in sSFR distributions in Figure \ref{fig:4p}, the $f_\mathrm{sf}$ of barred moderate-mass disk-dominated galaxies in interacting clusters ($0.585\pm0.068$) is far higher than that of non-barred ones ($0.387\pm0.046$) in the same clusters.

As shown in Figure \ref{fig:4p}, the $f_\mathrm{sf}$ in barred moderate-mass disk-dominated galaxies in non-interacting clusters ($0.400\pm0.043$) is slightly higher (but not statistically significant as in the case of the interacting clusters) than that of non-barred ones in non-interacting clusters ($0.328\pm0.022$). This could be also attributed to the bar-driven enhancement of star formation. However, in non-interacting clusters, the bar formation is likely not due to a cluster-wide mechanism occurring in a narrow time period as in the case of interacting clusters. Therefore, a number of bars formed recently (and subsequent star-formation activities triggered recently) would be smaller in non-interacting clusters than in interacting clusters. Accordingly, the $f_\mathrm{sf}$ value in barred moderate-mass disk-dominated galaxies is not as high as that of the counterparts in interacting clusters, since cluster environments are disadvantageous for preservation of star-formation activities.

According to our results, the ages of bar structures in interacting clusters are expected to be statistically younger than their counterparts in non-interacting clusters. Future studies on the ages of bar structures in interacting and non-interacting clusters should be able to verify this fact.
\\

\section{Summary} \label{sec:summary}

The first aim of this study is to investigate whether $f_\mathrm{sf}$ in interacting clusters is enhanced compared with that of non-interacting clusters. The second aim is to examine the link between the $f_\mathrm{sf}$ enhancement and the bar fraction enhancement in interacting clusters. To do so, we used the samples of galaxies and clusters from \citetalias{Yoon2019}, which are based on the MPA--JHU catalog. In total, 105 galaxy clusters at $0.015<z<0.060$ were examined, among which 16 are interacting clusters. The main conclusions are summarized as follows.

\begin{enumerate}
\item $f_\mathrm{sf}$ is moderately enhanced in interacting clusters compared with non-interacting clusters: $f_\mathrm{sf}$ is $1.19\pm0.08$ times higher in interacting clusters than in non-interacting clusters.

\item The enhancement of $f_\mathrm{sf}$ in interacting clusters occurs only in moderate-mass disk-dominated galaxies ($B/T\le0.5$ and $\log(M_\mathrm{star}/M_{\odot}) < 10.4$). This can be attributed to the relatively abundant gas contents in those galaxies compared to high-mass or bulge-dominated ones.

\item The enhancement of $f_\mathrm{sf}$ in moderate-mass disk-dominated galaxies in interacting clusters is directly related to the increase of the number of barred galaxies, which implies a connection between the star-formation enhancement and the bar formation by cluster--cluster interactions. 

\item Our results can be well explained by a mechanism that induces bars and triggers subsequent star-formation through the newly induced bars in disk galaxies in interacting clusters. One plausible physical mechanism is the time-dependent tidal gravitational field during the cluster--cluster interaction \citep{Bekki1999}.

\end{enumerate}

Our results imply that the most energetic phenomenon in large-scale environments such as cluster--cluster interaction can induce bars \citepalias{Yoon2019} and star formation at the same time in cluster galaxies. The exact details of how the cluster--cluster interaction induces star formation and bars need to be understood through future simulation and observational studies with a larger sample.
\\

\acknowledgments
We thank the anonymous referee for constructive comments that helped improve the content of the paper.
This work was supported by the National Research Foundation of Korea (NRF) grant, No. 2017R1A3A3001362, funded by the Korea government (MSIP).
This work was supported by a KIAS Individual Grant PG076301 at Korea Institute for Advanced Study.

\end{document}